\documentclass[12pt]{article} %

\usepackage {amsfonts}

 \UseRawInputEncoding

\begin{document}

\title{What Represents Space-time?  And What Follows for Substantivalism \emph{vs.} Relationalism and Gravitational Energy?\footnote{Forthcoming in Antonio Vassallo (Ed.), \emph{The Foundations of Spacetime Physics: Philosophical Perspectives}, Routledge }}
\author{J. Brian Pitts\footnote{University of Lincoln, University of Cambridge, and University of South Carolina} }

\date{18 January 2022} 

\maketitle

\begin{abstract} The questions of what represents space-time in GR, the status of gravitational energy, the substantivalist-relationalist issue, and the (non-)exceptional status of gravity are interrelated.  If space-time has energy-momentum, then space-time is substantival.  Two extant ways to avoid the substantivalist conclusion deny that the energy-bearing metric is part of space-time or deny that gravitational energy exists. Feynman linked doubts about gravitational energy to GR-exceptionalism, as do Curiel and Duerr;  particle physics egalitarianism encourages realism about gravitational energy.  

In that spirit, this essay proposes a third  possible view about space-time, one involving a particle physics-inspired non-perturbative split that characterizes space-time with a constant background \emph{matrix} (not a metric tensor), a sort of vacuum value, thus avoiding the inference from gravitational energy to substantivalism.  On this proposal,  space-time is $\langle M, \eta \rangle$, where $\eta=diag(-1,1,1,1)$ is a spatio-temporally constant numerical signature matrix, a matrix already used in GR with spinors.  The gravitational potential, to which any gravitational energy can be ascribed, is $g_{\mu\nu}(x)-\eta$ (up to field redefinitions), an \emph{affine} geometric object with a tensorial Lie derivative and a vanishing covariant derivative.  This non-perturbative split permits strong fields, arbitrary coordinates, and arbitrary topology, and hence is pure GR by almost any standard.  This razor-thin background, unlike more familiar backgrounds (\emph{e.g.}, Rosen's flat metric tensor field, Rosenfeld and M{\o}ller's orthonormal tetrad, and Sorkin's background connection), involves no extra gauge freedom and so lacks their obscurities and carpet lump-moving.  

	After a discussion of Curiel's GR exceptionalist denial of the localizability of gravitational energy and his  rejection of  energy conservation, the two traditional objections to pseudotensors, coordinate dependence and nonuniqueness, are explored. Both objections are inconclusive and getting weaker.  A literal interpretation involving infinitely many energies corresponding by Noether's first theorem to the infinite symmetries of the \emph{action} (or laws) largely answers Schr\"{o}dinger's false-negative coordinate dependence problem. Bauer's false-positive objection has multiple answers.  Non-uniqueness might be handled by Nester \emph{et al}.'s finding physical meaning in multiplicity in relation to boundary conditions, by an optimal candidate, or by Bergmann's identifying the non-uniqueness and coordinate dependence ambiguities as one. \end{abstract}


\section{Four Inter-related Questions}

	This essay considers four inter-related questions that are not usually all explicitly considered simultaneously, but which have some subtle connections worth uncovering.  
\begin{itemize} 
   \item  What represents space-time in GR?
  \item  What is the status of gravitational energy(s) in GR?
  \item  Substantivalism \emph{vs.} relationism: is space-time a substance? 
  \item  Is gravity exceptional, or just another field?
\end{itemize}
The first and third questions have been staples in philosophy for decades.  Gravitational energy, controversial in physics since 1918, has become a fashionable topic in philosophy within the last decade.  The fourth question, GR exceptionalism \emph{vs.} particle physics egalitarianism, has rarely been discussed in philosophy or history (but see \cite{KaiserGRParticle,TenerifeProgressGravity,LambdaMPIWG}), and is not often discussed \emph{explicitly} in physics either, but disagreement on it lies near the core of rival research programs in quantum gravity \cite[chapter 6]{Rovelli,BrinkDeserSupergravity,SmolinTroublePhysics}.  Given that  particle physics egalitarians rarely have been interested in philosophy until recently---the slogan ``shut up and calculate'' and Weinberg's chapter ``Against Philosophy'' \cite{WeinbergDreams} come to mind---there might be some low-hanging fruit to pick in applying the egalitarian perspective in relation to these other issues.  Feynman linked skepticism about gravitational energy to GR exceptionalism \cite[pp. 219, 220]{Feynman}.  Curiel's recent work, discussed below, also suggests this connection from the opposite direction \cite{CurielStressSHPMP}.   Duerr also makes this connection \cite{DuerrAgainstFunctionalGravitationalEnergy}.

This essay proposes a particle  physics-inspired answer to the question of what is space-time in GR, relates that answer to (anti)realism about gravitational energy, and provides another way to make gravitational energy(s) safe for  relationism.  This last conclusion  coheres with some themes of the dynamical view of space-time of Brown and (sometimes) Pooley \cite{BrownPooleyNonentity,BrownPhysicalRelativity}.  Previously I have found support in particle physics for at least some of Brown's project \emph{vis-a-vis} space-time realist critiques \cite{FMS,PittsScalar,ScalarGravityPhil,BrownFest}.  The result seems to be a coherent package of views that  has not previously been seen clearly as a whole and in detail, and perhaps  illustrates the fruitfulness of overcoming GR \emph{vs.} particle physics divide---a divide that philosophers and historians have inherited, along with a  GR exceptionalist bias, largely by default (but see  \cite{MenonSuperspace,SalimkhaniSpin2}).



\section{What Represents Space-time in GR?}

The question of what represents space-time in GR has been a standard topic in the philosophy  since the 1980s.  One obvious answer  is  $\langle M, g_{\mu\nu} \rangle,$ the manifold with the metric. It is tempting, at least to philosophers, to think that one first has a manifold with identifiable points, and then lays a metric on top of it, a task that leaves freedom of choice.  That path quickly leads to the hole argument, however, and is  not required  \cite[p. 155]{EinsteinSpaceNotLeftBehind} \cite{PittsSpinor,MenonAlgebraic}. 
 (If the points have some sort of essence that restricts the process of laying down the metric \cite{MaudlinEssence}, that is another matter.) 
 Another influential answer to the question is the bare manifold $M$ \cite{HoleStory}. A bare manifold presumably does not get its point individuation using promissory notes to be cashed out in terms of fields ($g_{\mu\nu}$ or otherwise) to be defined upon it.  Point individuation appears to be primitive, and the hole argument lurks again.  It is easy to feel dissatisfied, but harder to find an alternative. Iftime and Stachel's proposal to define space-time by projection down from a bundle deserves consideration  \cite{StachelIftime}, as do old and new efforts to avoid individuals \cite{DasguptaBareNecessities}.

Gravitational energy plays a key role in a foundational influential paper  on the hole argument \cite{HoleStory}, motivating the classification of the metric as part of the contents of space-time, as akin to matter  (not, of course, in the technical GR sense of matter as non-gravitational).  
\begin{quote} 
The metric tensor now incorporates the gravitational field and thus,
like other physical fields, carries energy and momentum, whose density is
represented by the gravitational field stress-energy pseudo-tensor.\ldots [which] forces its classification as part of the contents of spacetime.   \cite{HoleStory}
\end{quote}
This is surprising and perhaps too violent, one might think:  should not one be able to think of space-time in the absence of gravity, perhaps as something flat or trivial and in any case not energy-bearing, while energy-bearing gravitational waves are ascribed to gravity and not to space-time---to contents and not to the container, as it were?  Earman and Norton have considered that idea. 
\begin{quote} 
We might consider dividing the metric into an
unperturbed background and a perturbing wave in the hope that the latter
alone can be classified as contained in spacetime. This move fails since there
is no non-arbitrary way of effecting this division of the metric. \cite{HoleStory}
\end{quote}
Thus the claim that no principled split can be made plays a crucial role in forcing all of the metric, not just some energetic perturbation from triviality (flatness?), out of the space-time category and into the material category for Earman and Norton. If a sufficiently principled split exists, then one might have an intermediate option between space-time as the manifold and space-time and the manifold with the metric.

The claimed  non-existence of a non-arbitrary split into background and perturbation  may suggest GR exceptionalist sympathies:  distaste for background structures, for things that arguably act without being acted upon, along with a strong inductive claim over the trajectory of scientific progress.  On the other hand, the background structures that one typically sees put forward as potentially formalisms to be interpreted (as opposed to mere mathematical conveniences), such as  a background metric  \cite{Rosen1} or an orthonormal basis  \cite{MollerRadiation} or even a background connection  \cite{SorkinConnection}, have the disadvantage of introducing mathematical structure over and above what $g_{\mu\nu}$ requires (whether  $g_{\mu\nu}$ is primitive or as derived), along with a new gauge `group' (perhaps a Brandt groupoid) denying \emph{quantitative} significance to (most of) this extra structure and (in the case of a background metric or connection) bearing a striking resemblance to the coordinate freedom of GR. 
 Such background structures were proposed to improve the status of gravitational energy in GR, but the Pickwickian character of the improvement, however, was not always adequately recognized:  the advertised tensorial (coordinate-covariant) quality of the result was achieved at the expense of a less advertised new, comparably bad gauge dependence.  What the right hand gives, the left takes away \cite{Band2,BrazilLocalize}. With gravitational energy, the lump in the carpet has only been moved from coordinate dependence to gauge dependence.  

This paper, following a trajectory \cite{EnergyGravity,ConverseHilbertian}, avoids such  traditional background structures and seeks  a  middle path. The background can be taken to be the constant scalar numerical matrix $\eta=_{df} diag(-1,1,1,1)$ (up to trivial notational variants such as $diag(1,-1,-1,-1)$ or $diag(-1,-1,-1,1$)).  This matrix encodes a sort of `vacuum' value for space-time geometry, what remains (it is proposed) when not only all the fields standardly regarded as material are abstracted away, but also the gravitational field (in a sense to be made clearer).  In special relativistic theories without gravity, or even with gravity as described by certain theories (such as massive scalar gravity \cite{PittsScalar,ScalarGravityPhil}), $diag(-1,1,1,1)$  encodes the space-time geometry  in the absence of gravity, as expressed in coordinates that are as good as any---I avoid the word ``preferred'' advisedly.
In  General Relativity this matrix  often appears in weak-field expansions of GR and, more relevantly,  in  particle physics literature on quantum gravity in the tradition of covariant perturbation theory.\footnote{That covariant perturbation theory as a means of quantization did not work due to nonrenormalizability (infinities not removable without an infinite number of additional parameters) is not very relevant.  Moreover, after its apparent demise in the 1970s, a possible revival in supergravity emerged \cite{BernFinite3,BernFinite4}. In any case  one gets a quite satisfactory effective field theory \cite{DonoghueEffective}. }  Neither the matrix's constancy nor its being a scalar obviously makes its being a physical field problematic, unless perhaps one objects to things that `act without being acted upon' \cite{FriedmanJones,PooleyEncyclopedia}.  This same matrix also standardly  appears  in  GR with spinors, where no one seems to find it objectionable or thinks that it radically changes the character of the theory relative to tensorial matter fields.  (In James L. Anderson's analysis of absolute objects as amended and extended by Thorne, Lee, and Lightman, $\eta$ is a confined object, not an absolute object \cite{TLL}: it does not constitute the basis of a faithful representation of the manifold mapping group, or perhaps one should say, of coordinate transformations.)  From these angles, $\eta$ as a background representing space-time in GR is neither very novel nor very objectionable.

From a particle physics egalitarian perspective, $\eta$ plays a somewhat similar role to what  $g_{\mu\nu}$ plays in a GR-exceptionalist view.  In the latter, one can imagine abstracting away everything material (\emph{i.e.}, everything besides $g_{\mu\nu}$), and then seek to `build' a Lagrangian   by  coupling matter to $g_{\mu\nu}$ so as to give a dynamics to space-time and to matter coupled to space-time ($g_{\mu\nu}$); the Lagrangian should be a scalar density (up to a coordinate divergence) under coordinate transformations to give tensorial Euler-Lagrange equations when all fields are varied.  For the particle physics egalitarian, the matter-like aspects of gravity (having energy-momentum, \emph{etc.}, also emphasized by Rovelli, no particle physics egalitarian  \cite[p. 193]{RovelliHalfWoods}) call for  abstracting away one more field after the GR-exceptionalist stops:  one abstracts away the gravitational potential out of $g_{\mu\nu}$, leaving only $\eta.$  $\eta,$ being numerically constant, does not individuate space-time points; one has multiplicity without identity.   (There might be an interesting resonance with the thin substantivalism of Dasgupta \cite{DasguptaBareNecessities} and the Ramsification entertained briefly by Maudlin \cite{MaudlinEssence}.)  
It might help to think of the world as akin to some device held together with screws, in which  almost every piece  (all but one or two, analogous to the matter fields as listed by the GR exceptionalist) is screwed into some base in multiple places (analogous to coupling to $g_{\mu\nu}$).  The particle physics egalitarian claims that this base (analogous to $g_{\mu\nu}$) is itself composite:  the gravitational potential $\gamma_{\mu\nu}$ is one more matter-like field that can be unscrewed from the true base, which (to continue the analogy) is the matrix $\eta=diag(-1,1,1,1).$  Detailed features of the gravitational potential explain why it couples to everything in a universal way---why the other parts/fields `screw into' both the gravitational potential and the base in the same fashion.   If one imagines (re)assembling the world (or in any case its laws) by devising a suitable Lagrangian density, one has the space-time background $\eta=diag(-1,1,1,1),$ the gravitational potential $\gamma_{\mu\nu}$, and the (non-gravitational) matter fields as ingredients and needs to construct a Lagrangian density invariant (up to a divergence) under coordinate transformations.  Why seek a Lagrangian density of \emph{that} sort given these ingredients?  The particle physics spin $2$ derivations of Einstein's equations (with citations in \cite{DuffCovariant,ConverseHilbertian,SalimkhaniSpin2}) provide rather compelling guidance from very weak premises about avoiding violent instability, locality, (at least) Poincar\'{e} invariance, and the empirical fact of the bending of light.

The proposal about the gravitational potential overlaps a bit with a footnote by Pooley, who also raises several important challenges to taking the idea seriously.  His  view is that the best candidate (such as it is) for being the ``gravitational field'' is the
\begin{quote}
 \emph{deviation of the metric from flatness:} 
$h_{ab}$, where $g_{ab} = \eta_{ab} + h_{ab}.$ That this split is not precisely defined and does not correspond to
anything fundamental in classical GR underscores the point that, in GR, talk of the ``gravitational field'' is at best unhelpful and at worst confused. The distinction between background geometry
and the graviton modes of the quantum field propagating against that geometry is fundamental
to perturbative string theory, but this is a feature that one might hope will not survive in a more
fundamental ``background-independent'' formulation. \cite[footnote 34, p. 539]{PooleyEncyclopedia}\footnote{The referee  provided a timely reminder of this passage. }
\end{quote}  
In what sense is this split not precisely defined? Two senses come to mind. One  sense, at least if $\eta_{ab}$ is intended to be a flat metric \emph{tensor} (not a numerical matrix), is the extra gauge freedom that arises in relating the two metrics \cite{Grishchuk}, a gauge freedom that takes over much of the interesting role that coordinate transformations play in orthodox GR, thus separating the freedom to use spherical coordinates from notions of inertia and the like.  Moving the lump in the carpet does not flatten the carpet, however; hence I propose not a flat metric tensor, but a numerical matrix $diag(-1,1,1,1),$ leaving the coordinate freedom of GR to  play its classic role.  A second sense in which the deviation of the metric from flatness is not precisely defined involves field redefinitions.  Why should we take the gravitational field (though I use the term ``potential'') to be  the deviation $g_{ab} - \eta_{ab},$ and not $g^{ab} -\eta^{ab}$ (suppressing factors involving Newton's constant)?  There is no compelling reason.  Indeed the freedom to define the gravitational potential in different ways, many of them rather exotic (such as arbitrary powers of the metric, its inverse, or densitized relatives thereof---and beyond!) has been used as a resource to derive both Einstein's equations and infinitely many massive variants thereof \cite{OPSpinor,FMS,MassiveGravity1,MassiveGravity3} and to render the GR Lagrangian polynomial using either $(-g)^{\frac{5}{18}}g^{\mu\nu} - (-\eta)^{\frac{5}{18}}   \eta^{\mu\nu} $ or   $(-g)^{-\frac{5}{22}}g_{\mu\nu} - (-\eta)^{-\frac{5}{22}}   \eta_{\mu\nu}$  \cite{DeWitt67b}.  Ambiguity, however, is also present to some degree in geometrical GR.  Why should we take the field variable to the metric $g_{\mu\nu},$\footnote{The abstract index notation, often associated with lower-case Latin letters for four-dimensional quantities, has been claimed to have ``all the advantages of the component notation'' while avoiding  the disadvantage of obscuring the distinction between tensorial equations and equations for their components in some particular basis (perhaps adapted to the symmetries of some problem) \cite[p. 24]{Wald}. But what of expressive adequacy? Non-integral weight densities in the abstract index notation appear to be an unresolved problem---to say nothing of nonintegral powers of the metric \cite{OP}---which in turn tends to restrict what can be thought. } rather than $g^{\mu\nu}$ (as one occasionally sees), or more plausibly $\mathfrak{g}^{\mu\nu}$ ($= \sqrt{-g} g^{\mu\nu}$), which has pride of place in writing wave equations \cite{Papapetrou,ChoquetBruhatCauchyProblemGItCR} and simplifies the Lagrangian \cite{GoldbergConservation}?  Hence the problem of field redefinitions arises also in GR.  Admittedly the various choices seem more natural and more closely related through natural matrix operations such as inverse and determinant, as opposed to  power series expansions of (say) $(-g)^{\frac{1}{2}}g^{\mu\nu} - (-\eta)^{\frac{1}{2}}   \eta^{\mu\nu} $ in terms of $g_{\mu\nu} -   \eta_{\mu\nu}$ or \emph{vice versa}. (I use bimetric notation, but the simplification to $\eta=diag(-1,1,1,1)$ is obvious.) 
Viewing field redefinitions as choices of coordinates on a fiber might be an elegant approach for either the geometrical or the perturbative approach.  
 The geometrical view also might render more natural the Riemannian signature of $g_{\mu\nu}$ as a law, though perhaps nuts-and-bolts (quantum?) physics would prevent degeneracy or signature change from a particle physics standpoint.   On the other hand, it remains unclear on what grounds one should or even can postulate `equal-time' or space-like commutation relations without a background notion of causality \cite{NullCones1}.  
Finally, while the gravitational potential is uniquely defined only up to field redefinitions, the background matrix $\eta$ is unique (up to a conventional overall sign and a conventional choice to put `time', \emph{i.e.}, the sign that differs from the other three, first or last). 
Hence the ambiguity in defining the gravitational  potential does not infect space-time.

While one might see the appeal of a background of constant curvature (thus admitting a cosmological constant $\Lambda$), it might seem needlessly restrictive to specify a \emph{flat} background.  A background of constant curvature, however, apparently requires either explicit dependence on space-time coordinates (unlike $diag(-1,1,1,1,)$), which seems arbitrary, patchy (not attractive globally), and at best ugly, or  the use of tensor calculus with a background metric \emph{tensor} and consequent duplication of the gauge freedom.  Is representing space-time with the matrix $diag(-1,1,1,1)$ hence too fragile?  Perhaps it isn't, because Gia Dvali and collaborators argue that the need to accommodate theories requiring $S$-matrix formulations, including string theory, mandates a \emph{flat} background, not  one of constant (positive) curvature \cite{DvaliPositiveLambdaExclusion,DvaliSmatrixNoDeSitter}.  Hence there are modern independent plausible physical reasons for thinking that flat space-time plays a fundamental role. If so, why shouldn't gravitational energy refer to it somehow?

%


\section{Space-time Energy Implies Substantivalism?}

Plausibly, if  space-time has energy-momentum, space-time is substantival \cite{HoleStory,Hoefer}.  This claim has been shared by Earman and Norton, who affirm gravitational energy's reality  (but not localizability) and  seem \emph{prima facie} attracted to  substantivalism (while encountering objections to it), and Hoefer, who denies gravitational energy and sees relationism-friendliness as an important benefit.  
 Earman \& Norton  deny that $g_{\mu\nu}$ is part of space-time to avoid space-time energy.  
\begin{quote}
[GR] 
 has made most compelling the identification of the bare manifold with spacetime. For in that theory geometric structures, such as the metric tensor, are clearly physical fields in spacetime. [footnote suppressed]  The metric tensor now incorporates the gravitational field and thus, like other physical fields, carries energy and momentum\ldots 
 in a way that forces its classification as part of the contents of spacetime. 
 \cite{HoleStory}  
\end{quote} 
In partial contrast, Hoefer denies gravitational energy and saves relationism: 
\begin{quote}  
But if $T^{ab}=0$ `empty space' can carry \emph{genuine} energy-\\momentum of the gravitational field, then it (the empty space) should be counted as real also, and spacetime itself as represented by $g^{ab}$ should be considered substantial and real. \cite{Hoefer}  \\
If empty spacetime need not be thought to possess genuine energy, at least one
reason for considering it to be a substance is deflected. \cite{Hoefer} 
\end{quote}
While granting the  inference from space-time energy to substantivalism, the view proposed here affirms gravitational energy, while denying that gravitational energy is space-time energy.  Thus one finds another way to avoid an argument for substantivalism from gravitational energy.   One should consider, however, whether $\eta$ savors of absolute space(time) before settling on a verdict about substantivalism (\emph{c.f.} \cite{BrownPooleyNonentity}).

It should be noted that the inference from space-time  energy to  space-time substantivalism is intended to flow in only one direction.  While authors of diverse views (noted above) seem to accept that space-time energy is or would be sufficient for  space-time realism, there is no claim of a necessary condition.  Duerr articulates a view in which space-time is real without space-time energy, because for Duerr (like Hoefer), gravitational energy does not exist \cite{DuerrThesis}.


\section{Particle Physics Egalitarianism \emph{vs.} GR Exceptionalism} 

Partly  by default the philosophy of space \& time since the late 1970s has leaned toward GR exceptionalism.  Given the indifference (at best) of many particle physicists to philosophy, the only physicists who seemed relevant must have been general relativists.  But the very community identification of general relativists---the  sign on the door  indicates that one is attached to a particular theory of a particular force, akin to identifying with Weberian electrodynamics---reflects and inculcates GR exceptionalism.  GR exceptionalism \emph{vs.} particle physics egalitarianism is rarely explicitly debated (but see \cite{Feynman,DuffCovariant,KaiserGRParticle,Rovelli,SmolinTroublePhysics,BrinkDeserSupergravity,TenerifeProgressGravity,LambdaMPIWG}), but is clearly implicit in decades-long research programs in quantum gravity.  Thus canonical quantum gravity was not merely intended by GR exceptionalists to use Hamiltonian methods to merge quantum mechanics and GR, but also to leave ample room for revolutionary and nonperturbative consequences not expected or even accommodated using particle physicists' perturbative techniques \cite{SalisburySyracuse1949to1962}.  This section aims both to illustrate the two attitudes and to indicate  how the  particle physics egalitarian attitude might affect discussions of gravitational energy.  

	 Feynman taught a course at CalTech in 1962-3 in which he approached gravity as just another physical field, assumed like the others \emph{a priori} except where empirical facts implied a distinction.  
{\begin{quote} 
\ldots [M]eson theorists \ldots 
have gotten used to the idea of fields, so that it is not hard for them to conceive that
the universe is made up of twenty-nine or thirty-one other fields all in one grand
equation; the phenomena of gravitation add another such field to the pot, it is a
new field which was left out of previous considerations, and it is only one of the
thirty or so; explaining gravitation therefore amounts to explaining three percent
of the total number of known fields.
\cite[p. 2]{Feynman} \end{quote}  
Feynman, like various other authors \cite{Gupta,Kraichnan,Weinberg64d,Deser,SliBimGRG,ConverseHilbertian}, showed how to derive Einstein's equations with considerable rigor from very weak, plausible field theoretic postulates and a few empirical facts (such as the bending of light).  The rigor greatly exceeds, say, Einstein's arguments about rotation and misconceptions about conservation laws in his process of discovery \cite{EinsteinEnergyStability}.  One might expect that many physicists over decades would do better than one physicist in a few years having to invent many of the relevant tools.  But the ``spin 2'' derivations of Einstein's equations still seem not to get the attention that they deserve in philosophy, at least until recently \cite{SalimkhaniSpin2}. It was noticed recently that the spin $2$ derivations make crucial use of a form of Noether's converse Hilbertian assertion, that the energy-momentum complex's being the sum of a term vanishing with the field equations and a term with automatically vanishing coordinate divergence implies general covariance \cite{ConverseHilbertian}.

Feynman explained later how his particle physics egalitarian views affected the  debate on gravitational energy.  
\begin{quote} 
What is the power radiated by such a [gravitational] wave? There are a great many
people who worry needlessly at this question, because of a perennial prejudice
that gravitation is somehow mysterious and different---they feel
that it might be that gravity waves carry no energy at all. We can
definitely show that they can indeed heat up a wall, so there is no question 
as to their energy content. The situation is exactly analogous to 
electrodynamics\ldots. 
\cite[pp. 219, 220]{Feynman}
\end{quote} 
Feynman seems never to have given any very definite response to the usual worries about gravitational energy, but he seems committed  to the existence of gravitational energy and to  quasi-localization \emph{avant la lettre}:  heating up a wall implies that energy can be localized into small regions.  

Curiel's recent work underlines the connection between GR exceptionalism and doubts about gravitational energy.  As will be discussed below, his GR exceptionalism motivates giving up on the idea of conservation altogether \cite{CurielStressSHPMP}.  On the other hand, one might reconcile realism about gravitational energy with general relativist sympathies following  Rovelli: 
\begin{quote} 
 A strong burst of gravitational waves could
come from the sky and knock down the rock of Gibraltar, precisely as a
strong burst of electromagnetic radiation could. Why is the first ``matter''
and the second ``space''? Why should we regard the second burst
as ontologically different from the second?  \cite[p. 193]{RovelliHalfWoods} 
\end{quote} 
His book elaborates on matter-like aspects of gravity   \cite[p. 77]{RovelliBook}.

%


 \section{Three Questions about Gravitational Energy and a Neglected Option?} 

	Recalling the views of Earman \& Norton and of Hoefer on gravitational energy, one might consider three questions.  First, is gravitational energy real? Second, should the (formal) gravitational (pseudo-?)energy be attributed just to $g_{\mu\nu}$, not to something else in addition or instead? Third, is  $g_{\mu\nu}$ part of what represents space-time?  Four interesting sets of answers (perhaps among others) are: 
\begin{itemize} 
  \item  Gravitational energy is real and attributable just to $g_{\mu\nu}$, which is  part of space-time.  (Yes, yes, and yes) 
       \item  Gravitational energy is real and  attributable just to $g_{\mu\nu}$, but $g_{\mu\nu}$ is not part of space-time; only $M$ is \cite{HoleStory}. (Yes, yes and no)
     \item  Gravitational energy is not real, but gravitational pseudo-energy is  attributable just to  $g_{\mu\nu}$,  which is part of space-time  \cite{Hoefer}.  (No, yes and yes) 
  \item Gravitational energy is real, but it is attributable to the gravitational field and not to $g_{\mu\nu}.$  $g_{\mu\nu}$ is a composite entity made partly of non-energetic  space-time ingredient(s)  and partly of the physical-material gravitational field. (Yes, no and no) 
    \end{itemize}

	Something like  the fourth view was contemplated briefly by Earman \& Norton; has the view been set aside deservedly? Let us recall: 
\begin{quote} We might consider dividing the metric into an unperturbed background and a perturbing wave in the hope that the latter alone can be classified as contained in spacetime. This move fails since there is no non-arbitrary way of effecting this division of the metric. \cite{HoleStory}  \end{quote}
While they do not provide an argument there, arguments had been given before and are provided elsewhere by Norton.  A flat background metric introduces a whole  new gauge `group' changing the flat background tensor while leaving the effective curved metric alone \cite{Band2,Grishchuk,NortonConvention,BrazilLocalize,PetrovPitts}.  Even specifying a \emph{flat} background metric isn't nearly enough to remove arbitrariness.  But this paper proposes not a background metric tensor, but a scalar matrix $diag(-1,1,1,1)$, the same at every point in every coordinate system---the matrix to which one could reduce the components of a flat background metric by adapting coordinates.  The matrix $diag(-1,1,1,1),$ however, is just $0$s, $1$s, and a $-1$.  Absent a cosmological constant, this split  crops up in weak field problems in GR and in perturbative treatments of gravity more generally, classical or quantum.  This razor-thin background  provides most  benefits of a background metric or connection, with hardly any of the disadvantages. One has pure GR, or something extremely close.  
One can make sense of the background and the gravitational potential by recalling some lesser-known parts of the classical theory of geometric objects.


\section{From Tensors to Geometric Objects} 

Differential geometry  generalized from tensors to ``geometric objects'' \cite{Nijenhuis,Schouten,Yano,AczelGolab,Anderson,FriedmanAbsolute,FriedmanJones,PittsSpinor,ReadGeometricObjects}, a literature that made interesting progress into the 1960s.  While more general notions existed, the basic idea was to generalize the tensor transformation law to any local algebraic transformation law built using derivatives of one coordinate system with respect to another.  The transformation rule could involve higher derivatives (seen already with Christoffel symbols), be affine rather than linear (also seen with Christoffel symbols), or even be nonlinear. Approximately zero interesting examples of nonlinear or even affine geometric objects (except the connection or its projective and trace/volume pieces)  were presented, however.  That was historically contingent,  because particle physicists were reinventing  largely the same ideas at the same time (affine and nonlinear group realizations) and applying them to spinors in space-time \cite{DeWittDissertation,DeWittSpinor1950,OP,OPSpinor,PittsSpinor}. For present purposes, the relevant post-tensorial geometric objects are affine with only first derivatives. 

I suggest that the main culprit with the background structures traditionally used, is not the flatness or the background character, but the tensor character.  How about a \emph{constant scalar} background matrix $ \eta=diag(-1,1,1,1)$?  This quantity has been common  in  covariant perturbation theory \cite{GuptaPPSL2,Feynman63,OP,OPSpinor,Veltman}.\footnote{Actually much of that work used for the background matrix  $I=diag(1,1,1,1)$ and $x^4 = ict$, which is even more striking, but  a bridge too far, restricting the coordinates or admitting only a clumsy generalization.}  
To consider a  scalar background matrix in differential geometry, one can explore  transformation rules beyond $O^{\prime} = f(\frac{\partial x^{\mu^{\prime}} }{\partial x^{\nu}}) O$ \cite{Tashiro1,Nijenhuis,Yano,AczelGolab,SzybiakCovariant}. Affine geometric objects have some nice properties:  non-tensorial behavior gets excised various contexts.  The tensorial Lie derivative of a connection is a somewhat familiar but  complicated example with second derivatives.  A  less familiar  example is a tensor (or tensor density) minus  some  constant(s), perhaps such as a `vacuum' value.  Splitting $g_{\mu\nu}$  into a background matrix $\eta$ and a gravitational potential $\gamma_{\mu\nu}$  (or building it from them, depending on what one takes to be primitive) provides a physically interesting example:  the gravitational potential $\gamma_{\mu\nu}$.  One  could as easily put the indices up as down, and/or give  any density weight,  integral or not \cite{OP,MassiveGravity3}. 


%
%

Assuming a metric $g_{\mu\nu},$ we can define a gravitational potential (or ``perturbation,'' though there is {no} assumption of smallness) by subtracting  $\eta$, a sort of vacuum value, not necessarily in the QFT sense, but the default value when nothing interesting is happening and descriptive simplicity is employed.  This procedure is routine in testing for stability, whether in the Higgs mechanism or in applied mathematics  \cite{KhazinShnolBook}. Choosing the normalization to give the Lagrangian standard (non-geometric) dimensionality and depend on the gravitational potential in the standard way, one has  $\gamma_{\mu\nu} =  (g_{\mu\nu} - \eta)/\sqrt{32 \pi G}$ (or the like under  algebraic field redefinitions involving $\mathfrak{g}^{\mu\nu} - \eta$, \emph{etc.}, reasonable choices agreeing in their lowest order traceless parts in approximately Cartesian coordinates given suitable normalization \cite{OP}). I (usually)  avoid writing  $\eta$ as $\eta_{\mu\nu}$ to avoid the notational suggestion that $\eta$ is a tensor; it is  a matrix-valued scalar.   %

Before delving further into the technicalities, one can  avert some tempting misconceptions.  First, this definition is non-perturbative because  $\gamma_{\mu\nu}$ is \emph{not assumed small}; neither the physics nor the coordinates are assumed  `mild'; one could use spherical  coordinates falling into a black hole or use approximately Cartesian coordinates in a `wrong' order such as $\langle x,t,y,z \rangle$ (albeit with a large effect on the gravitational potential value).  No series expansion or restriction  to achieve convergence is used.  Second, unlike a flat background metric tensor, the matrix $\eta$ imposes no topological restrictions.  It is merely part of a change of variables: one takes the components of the metric tensor (or some similar quantity) and subtracts $-1,$ $1$ or $0$.  Third, the coordinates are completely arbitrary; no  gauge fixing or coordinate condition, whether involving equations or inequalities (such as for time for a Hamiltonian) is employed.   Fourth, the success or failure of quantum gravity programs that made use of this expansion perturbatively is irrelevant.  The use is not intrinsically quantum, and one could quantize canonically or with a path integral or perhaps on some other way.  This is pure GR, or within a hair's breadth thereof, with an innocent change of variables.  Fifth, there is no extra gauge freedom,  only the traditional coordinate freedom of GR. Nothing, or nearly nothing extra has been introduced, and no new gauge freedom to deny its physical meaning is present or needed.   $\eta$ appears in GR with spinors anyway and so cannot be very bad.  There is little or no ontology for this razor-thin background. These remarks and those above are doubtless not the last word needed, but they might prime the pump.

With such misconceptions averted, we can explore the classical differential geometry of affine geometric objects. From $\gamma_{\mu\nu} =  (g_{\mu\nu} - \eta)/\sqrt{32 \pi G}$ and the tensor transformation rule   $$g_{\mu\nu}^{\prime} = g_{\alpha\beta}  \frac{\partial x^{\alpha} }{\partial x^{\mu^{\prime} }} \frac{\partial x^{\beta} }{\partial x^{\nu^{\prime} }} $$ one has: 
$$\gamma_{\mu\nu}^{\prime} = \gamma_{\alpha\beta}  \frac{\partial x^{\alpha} }{\partial x^{\mu^{\prime} }} \frac{\partial x^{\beta} }{\partial x^{\nu^{\prime} }}  +  \frac{ \eta_{\alpha\beta}}{\sqrt{32 \pi G} }  \left(   \frac{\partial x^{\alpha} }{\partial x^{\mu^{\prime} }} \frac{\partial x^{\beta} }{\partial x^{\nu^{\prime} }}  - \delta^{\alpha}_{\mu} \delta^{\beta}_{\nu} \right). $$ This is a local geometric object, with components at every point  in every coordinate system covering that point and a  transformation rule relating any two coordinate systems covering that point.  It is an affine, first-differential-order geometric object, falling in between tensors and connections.  Affine geometric objects have \emph{linearly} transforming Lie derivatives \cite{Tashiro1,Nijenhuis,Yano}:  the non-tensorial part of the transformation rule is shaven off, making the Lie derivative a tensor:  $$\pounds_{\xi} \gamma_{\mu\nu} = \xi^{\alpha} \gamma_{\mu\nu},_{\alpha} + \gamma_{\alpha\nu} \xi^{\alpha},_{\mu} + \gamma_{\mu\alpha} \xi^{\alpha},_{\nu} + \frac{\eta_{\alpha\nu} \xi^{\alpha},_{\mu} + \eta_{\alpha\mu} \xi^{\alpha},_{\nu}}{ \sqrt{32\pi G} }= \frac{1}{\sqrt{32 \pi G}} \pounds_{\xi} g_{\mu\nu}.$$
Thus symmetries of the metric (Killing vector fields) are readily expressed as symmetries of the gravitational potential.

What of covariant differentiation? There is a little known  formula  for the covariant derivative of any first-order (even nonlinear) geometric object $\omega^N$ \cite{SzybiakCovariant}:
 $$ \nabla_{\alpha} \omega^N = \partial_{\alpha} \omega^N + \Gamma^{\kappa}_{\alpha\lambda} \lim_{\frac{\partial x^{\prime}}{\partial x} \rightarrow I}  \frac{\partial \omega^{N^{\prime}}}{\partial \left(\frac{\partial x^{\kappa^{\prime}}}{\partial x^{\lambda}}\right) }.  $$
Thus  $\nabla_{\alpha} \gamma_{\mu\nu} = \gamma_{\mu\nu},_{\alpha} -\frac{1}{32 \pi G} (\Gamma^{\kappa}_{\alpha\nu} g_{\mu\kappa} +  \Gamma^{\kappa}_{\alpha\mu}  g_{\kappa\nu}),$ which  agrees with the metric's covariant derivative up to a constant factor: $\nabla_{\alpha} \gamma_{\mu\nu} =  \frac{1}{\sqrt{32 \pi G}}   \nabla_{\alpha} g_{\mu\nu}.$ 
If the covariant derivative is taken using a  $g_{\mu\nu}$-compatible connection $\Gamma^{\kappa}_{\alpha\lambda},$  the result vanishes.  Thus $\gamma_{\mu\nu}$'s Lie and covariant derivatives stand in for $g_{\mu\nu}$'s, as expected because  $\Gamma^{\kappa}_{\alpha\lambda}$ can be built from $\gamma_{\mu\nu}$ and $\eta,$ and any  derivative of $\eta$ vanishes.  One could  do all of Riemannian geometry in terms of $\gamma_{\mu\nu}$ and $\eta.$ Likewise one could derive the Euler-Lagrange equations and the gravitational energy-momentum pseudotensor(s) for GR in terms of  $\gamma_{\mu\nu}$ and $\eta.$

The matrix $\eta$ is no newcomer to gravitational energy, because both modern and classical works on gravitational energy have in some cases made use of a reference configuration.  Such modern work includes Nester and collaborators  \cite{ChangNesterChen,NesterQuasi} and overlapping teams including Grishchuk, Petrov, Katz, Bi\v{c}\'{a}k and  Lynden-Bell \cite{Grishchuk,KatzBicakLB,PetrovKatz2}.  One then faces the question of the gauge freedom in relating the two metrics. On the other hand, the background matrix  $diag(-1,1,1,1)$ is unique up to conventional choices (actually $diag(-1,-1,-1,1)$) in the  Papapetrou-Belinfante pseudotensor \cite{Papapetrou}.\footnote{Presumably no one will think that the matrix $(diag(-1,1,r^2, r^2 sin^2 \theta),$ for example,  is a plausible candidate to appear in laws of nature, although it is also, like $diag(-1,1,1,1),$ a matrix of components of a flat metric tensor in certain coordinates.}   Papapetrou's interpretive concerns about his pseudotensor formalism, I suggest, are largely addressed \emph{via} affine geometric objects (addressing his concerns about a ``system of some `auxiliary numbers' '') and the multiplicity of gravitational energies (addressing gauge dependence) \cite{EnergyGravity}.  Papapetrou's eventual introduction of a flat metric \emph{tensor} instead of a numerical matrix---the opposite of the move that I suggest---motivated him to gauge-fix the relation between the two metrics in order to avoid an infinity of distinct results, an infinity strikingly resembling coordinate dependence.  But his gauge fixing differs only formally from Fock-style fixation of the coordinates with the harmonic condition in a single-metric formalism \cite{Fock}.  If gauge-fixing is a satisfactory solution, then why not just fix harmonic coordinates in GR and declare gravitational energy to be localized in the true (\emph{e.g.}, harmonic) coordinates?  
 It seems to me advantageous to accommodate $diag(-1,1,1,1)$ and the metric perturbation within differential geometry and keep the coordinate freedom as it was.


\section{What Represents Space-Time:  A Proposal}

Now one can make more sense of the particle physics-inspired proposal for what represents space-time.  In between the standard suggestions of  $\langle M, g_{\mu\nu} \rangle$ and just $M$, is the proposal:  space-time is $\langle M, \eta \rangle.$  This suggestion might ring a bell by now:  ``We might consider dividing the metric into an unperturbed background and a perturbing wave in the hope that the latter alone can be classified as contained in spacetime.''   \cite{HoleStory}   I suggest that a  non-arbitrary  division is  $g_{\mu\nu} = \eta + \sqrt{32 \pi G} \gamma_{\mu\nu}$  (\emph{mutatis mutandis} with field redefinitions for the gravitational potential); hence $\eta$ is the unperturbed background and $\gamma_{\mu\nu}$ (or some equivalent entity) is the perturbing `wave' (not necessarily wavy) contained in space-time. Is this the golden mean, the Goldilocks zone? On this proposal,  \emph{much of} the metric is part of space-time, as one might prefer. If nothing much is happening (weak fields), and coordinates are adapted to this situation (not far from Cartesian and with time in  the temporal slot \cite{BilyalovSpinors,PittsSpinor}), then the value of $g_{\mu\nu}$ is approximately $\eta.$  On the other hand, $\eta$ savors of  Minkowski space-time, which by some lights is a relationism-friendly ``glorious non-entity''  \cite{BrownPooleyNonentity}, though one might also consider a resemblance to traditional absolute space(time). Gravitational energy is not due to space-time (though  $\eta$  might appear in it), but due to the gravitational potential $\gamma_{\mu\nu},$ likely quadratic in its first derivatives and possibly having some second derivatives (especially spatial or mixed).  If gravitational energy is not space-time energy, then one can believe in gravitational energy, even in infinitely many localized gravitational energies, with no pressure toward affirming substantivalism from such energy.  With gravitational energy ascribed to $\gamma_{\mu\nu}$, not  space-time, one can  do justice to the sensibility that the $g_{\mu\nu}$ is in many ways like a matter field  \cite[p. 77]{RovelliBook} \cite[chapter 9]{BrownPhysicalRelativity}---an idea linked to Rovelli's gravitational energy realism  \cite{RovelliHalfWoods}.  As Hoefer said, ``[i]f empty spacetime need not be thought to possess genuine energy, at least one reason for considering it to be a substance is deflected.''  \cite{Hoefer} Now this conclusion can be reached in another way, accepting gravitational energy but denying space-time energy.


\section{Curiel on  Gravitational Energy} 

The idea of  gravitational energy  has received philosophical attention recently from the GR exceptionalist perspective as well.  Curiel has underscored the non-existence of a local gravitational stress-energy tensor given certain assumptions and has suggested giving up  the  usual idea of conservation in favor of fungibility.  
\begin{quote} 
I prove that, under certain natural conditions, there can be no tensor whose interpretation could be
that it represents gravitational stress-energy in general relativity. It
follows that gravitational stress-energy, such as it is in general
relativity, is necessarily non-local. \cite{CurielStressSHPMP}  \end{quote} 
For Curiel the inference from non-tensoriality (or perhaps more generously, not being a geometric object) to non-localizability is rather direct.  A local but non-tensorial object, such as the Noether operator \cite{SchutzSorkin,SorkinStress} or the closely related and more pedestrian pseudotensor(s), is not a candidate because it is not a physical quantity \cite[p. 97]{CurielStressSHPMP}.

{The fact that integrals over pseudotensors depend only on coordinates at the boundary \cite{NesterQuasiPseudo}  indicates that the  temperature in Curiel's coffee is less coordinate-dependent than one might have feared.  His concern about using a pseudotensor to ascertain how much a gravitational wave partly absorbed by a  piezoelectric stick would warm his coffee, is unclear to me.   If the concern is that it can be used but would give different answers in different coordinate systems, that seems unlikely   because physical objects, such as coffee and thermometers, are  not sensitive to a choice of coordinates. The piezoelectric stick does not need to ```know' which of Pitts's `localized energies' it should draw on'' because any of them will do.  Recalling that a pseudotensor conservation law is logically equivalent to Einstein's equations \cite{Anderson,EnergyGravity}, the conservation law is  a way of expressing the content of GR  that shows that a sum of material energy and gravitational (pseudo-?)energy satisfies the continuity equation.  

Curiel has not specified the wavelength of the gravitational wave in question, but the ratio of this wavelength to the cup-detector affects the analysis  \cite{SchutzRicciGravitationalWaves}. For wavelengths small compared to the detector, the problem is largely  included in Schutz and Ricci's treatment.  After discussing how one can treat gravitational waves on a gently varying background as a type of matter and use Isaacson's averaging over a few wavelengths to get a localization to that scale, they comment:
\begin{quote}
 In the textbooks you will find discussions of pseudotensors,\ldots  of Noether theorems and formulas for energy, and so on.
None of these are worse than we have presented here, and in fact all of
them are now known to be consistent with one another, if one does not ask
them to do too much. In particular, if one wants only to localize the energy
of a gravitational wave to a region of the size of a wavelength, and if the
waves have short wavelength compared to the background curvature scale,
then pseudotensors will give the same energy as the one we have defined
here. 
\end{quote} 
Astrophysically plausible gravitational waves will tend to have wavelengths longer than a coffee cup, however, and a fundamental treatment ought to be able to handle all cases.   Hence not all of Curiel's question is  answered by Schutz and Ricci. My sense is that the question, if not already answered somewhere, is more like a puzzle than an anomaly (to borrow some Kuhnian concepts for an idea that hardly constitutes a paradigm sociologically):  the question is worth answering, and there is no good reason to doubt that it can be answered.} 

 Strikingly, symmetries of the action and Noether's theorem play no role for Curiel. Neither does the continuity equation appear, so Curiel does not entertain anything that could yield results such as  $E=constant$.  
This absence is not an oversight, but a principled inference from the premises adopted. 
\begin{quote} 
The formulation of the First Law [of thermodynamics] I rely on is somewhat
unorthodox: that all forms of stress-energy are in principle ultimately
fungible---any form of energy can in principle be transformed
into any other form [footnote suppressed]---not necessarily that there is some
absolute measure of the total energy contained in a system or set of
systems that is constant over time. \cite{CurielStressSHPMP}  
\end{quote} 
If coordinate-free GR doesn't have it, Curiel doesn't need it. Hence he does not need energy conservation, a view that some will consider bold, but a  coherent view that, used carefully,\footnote{Elsewhere I have discussed how the supposed absence of conservation laws in GR encourages  some people to object spuriously to the theory and  others to think that energy non-conserving processes are thereby licensed \cite{EnergyGravity,EnergyMental}. } will never yield a false  prediction (unless lower-brow approaches to GR would also).  

One might wonder, however, whether fungibility is an adequate version of, or substitute for, the First Law of Thermodynamics.  In the  multi-trillion-dollar foreign exchange market, of which the most obvious  manifestations at airports offer the clearest insights, one is given the opportunity to convert (say) USD to GBP at one rate, or the reverse at a very different rate. At airports this  ``spread'' is enormous, so one  could convert USD to GBP and back, leaving with  far fewer USD than one started with, however.   
Hence mere fungibility, with no chance of recovering one's starting point, is not clearly a version of the First Law.  Possibly it is something like a combination of the First and Second Laws at best: even if perhaps money is conserved,  your money tends to decrease.  But a clear analog of the First Law only is not to be found in mere fungibility without quantitative bookkeeping implying that the starting configuration is in some respects preserved.


 One might also question the mathematical basis for requiring a \emph{symmetric covariant} stress-energy tensor for gravity, in tension  with  what Lagrangian field theory offers.  For a scalar field, Noether methods offer for the stress-energy a  \emph{mixed} $(1,1)$ weight $1$ tensor \emph{density} $\mathfrak{T}^{\mu}_{\nu}$ to give  $\mathfrak{J}^{\mu}[\xi] = \mathfrak{T}^{\mu}_{\nu} \xi^{\nu}$. A displacement vector yields a conserved current that is a  tangent vector \emph{density} of weight $1$, thus such that   $\nabla_{\mu} \mathfrak{J}^{\mu} \equiv  \partial_{\mu} \mathfrak{J}^{\mu} =0 $.  Given that only a partial divergence  $\partial_{\mu} \mathfrak{J}^{\mu}=0$ has a chance of integration  to $E=constant$ \cite[pp. 236, 269-271]{WeylSTM} \cite[p. 280]{Landau} \cite[p. 465]{MTW} \cite[p. 139]{LordTensors} \cite[p. 141]{Stephani}, this is the ideal case, so requiring in advance something different (a symmetric stress-energy tensor) shows a lack of interest in getting conserved quantities, the obtaining of which usually have been considered a core part of the issue.  As noted earlier, conservation of angular momentum requires not symmetric stress-energy, but only rotational symmetry of the action, because angular momentum needn't be     $ x^{[\alpha}T^{\mu]\nu}$ 
\cite{BergmannThomson,ForgerRomerStress}; there can be a spin contribution. For vector matter, similar Noether methods give a current  $\mathfrak{J}^{\mu}[\xi]$ that is tensorial but not algebraic in $\xi^{\nu}$. Hence one cannot peel off the vector field $\xi^{\mu}$ and get a stress-energy tensor, as one can with a scalar field.  For GR, $\mathfrak{J}^{\mu}[\xi]$ either is non-tensorial or has higher derivatives, depending on which Lagrangian density one uses \cite{SorkinStress}. If one asks GR what the Noether mathematics means  instead of imposing requirements by hand, it (mostly?) makes sense:  the  Noether operator generalizing $\mathfrak{J}^{\mu}[\xi] ,$  depending  differentially on $\xi^{\mu},$  yields a pseudotensor if one  sets $\xi^{\mu} = (1,0,0,0)$. Abstaining from this sort of mathematics  to remain pure of coordinates leads to giving up conservation laws that do exist.   A criterion for physical quantities that excludes the Noether operator is not obviously appropriate.

It is also  obscure what the rules are in Curiel's admittedly utopian quest for a   gravitational stress-energy tensor  $S_{\mu\nu}$.  
This hypothetical entity is supposed to be symmetric ($S_{[\mu\nu]}=0$) and to have vanishing covariant divergence ($\nabla_{\mu} S^{\mu\nu}=0$). 
If that were true, then one could add it to the material stress-energy tensor as follows:  
\begin{eqnarray*}  \nabla_{\mu} (\sqrt{-g} S^{\mu\nu} + \sqrt{-g} T^{\mu\nu})  = \partial_{\mu} (\sqrt{-g} S^{\mu\nu} + \sqrt{-g} T^{\mu\nu})  + (\sqrt{-g} S^{\mu\alpha} + \sqrt{-g} T^{\mu\alpha}) \Gamma^{\nu}_{\mu\alpha} \nonumber \\ + (\sqrt{-g} S^{\alpha\nu} + \sqrt{-g} T^{\alpha\nu}) \Gamma^{\mu}_{\mu\alpha} - (\sqrt{-g} S^{\mu\nu} + \sqrt{-g} T^{\mu\nu}) \Gamma^{\alpha}_{\mu\alpha} 
\end{eqnarray*}
(using the covariant derivative of a density of weight $1$ for the last term \cite{Anderson})
\begin{eqnarray*}  = \partial_{\mu} (\sqrt{-g} S^{\mu\nu} + \sqrt{-g} T^{\mu\nu})  + (\sqrt{-g} S^{\mu\alpha} + \sqrt{-g} T^{\mu\alpha}) \Gamma^{\nu}_{\mu\alpha}=0.
\end{eqnarray*} 
This result is \emph{incompatible} with the  continuity equation $$\partial_{\mu} (\sqrt{-g} S^{\mu\nu} + \sqrt{-g} T^{\mu\nu}) \stackrel{?}{=}0$$ because  $ (S^{\mu\alpha} +  T^{\mu\alpha}) \Gamma^{\nu}_{\mu\alpha} \neq 0$ typically.  Hence the resulting equation would \emph{prohibit} the existence of a conserved energy $E=constant.$ What Curiel seems to regard as desirable given his heuristics but, alas,  impossible would in fact be  disastrous.  
Fortunately GR does permit $E=constant$ for asymptotically flat metrics because it admits the continuity equation, albeit for a quantity of which Curiel does not approve.  Thus  $S_{\mu\nu}$  wouldn't fit  GR.  Curiel agrees, but  for different reasons involving his proof, which depend on premises not following from Einstein's equations.  
\begin{quote} The existence of a gravitational stress-energy tensor, however, would necessarily entail that we modify our understanding and formulation of general relativity. That is why this argument is only \emph{ex hypothesi}, and not meant to be one that would make sense in general relativity as we actually know it. \cite{CurielStressSHPMP} \end{quote} 
If the discussion both assumes the supposed `spirit' of GR and contradicts GR's laws, it is unclear what the rules are or what proprietary question is being addressed instead of the usual gravitational energy localization problem in GR.  I am not convinced that one should heed guidance from such an antinomian spirit.  


\section{Two Standard Worries about Gravitational Energy}

I turn now to the two standard objections to pseudotensors and argue that both  have gotten weaker in the last 20+ years.  The two standard objections are that a  pseudotensor depend essentially on coordinates, which nothing physically real would  do, and that the pseudotensor is nonunique, which is also incompatible with physical reality.   Pseudotensors relate weirdly to coordinates in at least two ways:  false positives and false negatives.   Schr\"{o}dinger presented an early false-negative objection:  Einstein's gravitational energy-momentum pseudotensor vanishes outside a round heavy body in some coordinates, but surely there should be gravitational energy outside a round heavy body, and whether there is gravitational energy should not depend on coordinates  \cite{SchrodingerEnergy,Cattani}. Bauer presented a mirror-image objection that same year, a false positive: Minkowski space-time in (unimodular) spherical coordinates has a nonzero energy density and even infinite total energy  \cite{BauerEnergy,Cattani}. The nonuniqueness objection is that there are many comparably good candidates, and they cannot all be real, so plausibly none of them is real.   The coordinate dependence objection goes back to the 1910s, though there was interesting activity (such as by M{\o}ller among others) in the late 50s-early 60s, while the nonuniqueness objection seems to have become serious in the 1950s with the Landau-Lifshitz pseudotensor \cite{Landau}, Goldberg's infinity of pseudotensors \cite{GoldbergConservation}, and others.  Thus most people have long since given up on localization.  Some reject gravitational energy outright in light of its apparently inconsistent properties \cite{Hoefer,DuerrFantasticBeasts}.  This is a principled view that makes at least as much sense as the standard view \cite{MTW} that gravitational energy exists but is not localizable, a claim that Norton also finds obscure  \cite{NortonLearnOntology}.
  I argue that both worries are inconclusive and getting weaker recently. I also note Read's recent functionalist defense of gravitational energy \cite{ReadEnergy}.


%

%


 I have explained previously how asking Noether's first theorem how many conserved energies to expect---one for each rigid symmetry of the action, hence infinitely many \cite{BergmannConservation}---resolves Schr\"{o}dinger's false-negative objection \cite{EnergyGravity}.  Lacking a coordinate transformation is not a bug \cite{ReadGeometricObjects}.  In fact it is a feature: it permits the expression of infinitely many energies with only $10$ or $16$ components \cite{EnergyGravity}.

 To give an analogy, one might be puzzled by the inequivalence under translation (analogous to lack of a coordinate transformation rule) between  ``Mar\'{i}a   es alta'' (tall) and ``Mary is  short''---unless Mar\'{i}a  $\neq $ Mary, in which case there is no reason to expect equivalent heights. {If the comparative and context-dependent nature of ``tall'' and ``short'' are objectionable, then one can change the example to involve different unit systems.  Perhaps Mar\'{i}a is $n$ meters tall (in Spanish) and Mary is $x$ feet $y$ inches tall.   But most of us cannot make such conversions sufficiently accurately without calculation, so the contradiction is not evident. Inequality implies $12x+y \neq \frac{100 n}{2.54}.$ } 




 Bauer's false  positive objection can be criticized on various technical grounds.  First, it is unclear that unimodular spherical coordinates (or garden-variety spherical coordinates, for that matter) should be regarded as covering the whole manifold; by modern standards they don't.  Second,  such coordinates make  Einstein's $\Gamma-\Gamma$  action diverge. If the field equations and the canonical energy-momentum pseudotensor are derived from the action, and this coordinate system makes the action diverge, why admit them?  Third, there is a little-recognized nonuniqueness in the Lagrangian density which renders the Einstein pseudotensor optional as the canonical energy-momentum pseudotensor even  given metric variables (not to mention non-canonical pseudotensors).  In  Maxwell's electromagnetism,  one could include the contraction $(\partial_{\mu} A^{\mu})^2$ in the Lagrangian by adding a total divergence, and thus can write down a $1$-parameter family of equivalent Lagrangian densities differing by a total divergence.  GR admits a similar ambiguity at least linearly \cite[p. 647]{Ohanian}. This ambiguity seems to disappear in an exact treatment in metric variables, but it surfaces using an orthonormal basis, as in M{\o}ller's work.  If one wishes, one can  gauge-fix the tetrad into the symmetric square root of the metric  \cite{DeWittSpinor,MollerRadiation,OP,EnergyGravity,PittsSpinor}, which depends in a nonlinear way on the metric and on $\eta$. As long as one does not try to put a time coordinate in a spatial place or \emph{vice versa} (roughly) \cite{PittsSpinor}, one has an alternative GR Lagrangian built out of the metric and $\eta$ with only first derivatives.  Hence there is an apparently previously unrecognized $1$-parameter ambiguity of GR Lagrangians \emph{in metric variables}, leading to a similar $1$-parameter ambiguity of canonical energy-momentum complexes, including the (gauge-fixed) M{\o}ller tetrad complex.  But it gives $0$ energy for Minkowski space-time \cite{MollerRadiation}.  Thus it is unclear that Bauer has used admissible  coordinates, especially given his (Einstein $\Gamma-\Gamma$) action, and unclear that he has used the correct action, because an action exists that  avoids Bauer's false positive objection even in his unimodular spherical coordinates.  Recall that  M{\o}ller's tetrad energy-momentum expression  is tensorial (not a pseudotensor) under changes of coordinates, though dependent on the local Lorentz gauge. The local Lorentz gauge freedom can be fixed using the symmetric tetrad gauge condition to give a metric formalism with help from $\eta$ \cite{EnergyGravity}; some other ways of fixing the local Lorentz gauge give unreasonable results for the mass-energy \cite{MollerTetradEnergies}.    One  might also take the view, already familiar for spherical symmetry, that one should make use of symmetries when they exist \cite[p. 603]{MTW}.  A plausible generalization is that one should adapt one's coordinates as far as possible to the largest set of commuting Killing vector fields when it is unique \cite{EnergyGravity}.  Then  one should use Cartesian coordinates for Minkowski space-time, which would resolve Bauer's objection even given the Einstein pseudotensor.   Hence there are several  resolutions  of Bauer's objection.  

Pseudotensoriality occurs in ordinary life, as in the colors of flags.   In most places where English is the main language, ``the flag is red, white and blue'' is true.  This claim admits translation into a true statement in France, because the French flag has the same colors.  But it fails when translated into German or Spanish (at least in Spain) or most other languages.  Failure of translation here is not mysterious:  one is naturally referring to one's own national flag, so the statement in different languages has different referents (using the nation-state approximation to make the analogy vivid: one flag per language, which is more accurate in some places than others).   In advanced countries, no normal person older than perhaps $5$ years is unaware of the existence of other countries, so it is difficult not to know of the multiplicity of flags, so ``the'' flag will mean our flag.  But if one somehow managed not to know that, while knowing multiple languages---perhaps one is part of an educational experiment in a totalitarian country---then one might only know of one flag and think that there is only one flag.  Then one might read (perhaps due to a gap in censorship)  apparently plausible  but apparently incompatible statements such as ``Die Flagge ist schwarz, rot und gold'' and ``the flag is red, white, and blue.'' The paradox would be resolved by learning that there are different countries with different flags.  Flags and languages have a (more or less natural) relationship.\footnote{Obviously politically fraught real-world  issues are glossed over for the sake of the analogy.  No views are intended about independence movements, recognition of minority languages/dialects, \emph{etc.}  }  Perhaps energies and coordinate systems do as well, such as coordinates in which the corresponding displacement takes the form $(1,0,0,0)$ or the like.  

Above it was noted that pseudotensors are economical:  one is enabled to say infinitely many distinct things with a $10$- or $16$-component entity; the economy of pseudotensorial policies also has real-world examples, including the publishing industry and  some multilingual academic writing.  Whereas tensor calculus says the same old thing infinitely many times using the tensor transformation rule, no publisher feels obliged to publish  translations of all of its books in  every language. (`Coordinate-free' publication in terms of propositions or the language of thought, as opposed to publication in a language, is only  science fiction at this stage.) 
 In some academic fields, such as ancient or medieval philosophy, one finds  untranslated Greek or Latin text in  an article written in (usually) French, German or English, with along with untranslated quotations from the other two modern languages.  
While the audience able to appreciate such work fully is not large, it does exist.  Perhaps gravitational energy also involves a sort of multilingualism to accommodate the natural connections between energies and coordinate systems.


\section{Nonuniqueness Objection and 3 or 4 Possible  Answers} 

	One would expect the pseudotensor to be unique if it represents real gravitational energy.  But there are infinitely many candidates.  So the pseudotensor does not represent real gravitational energy.  This seems to be how the nonuniqueness objection runs.  There are, however,  four interesting replies to this objection.

An initial reply, which  appeals to the widespread acceptance of material  energy $T^{\mu\nu},$ is a  \emph{tu quoque} response \cite{EnergyGravity}:  gravitational energy is not qualitatively  worse off than supposedly unproblematic material energy.  Even a scalar field in flat space-time suffers from nonuniqueness due to a multiplicity of comparably plausible candidates, due to the ``improved'' energy-momentum  tensor, which has certain advantages  \cite{ImprovedEnergy}. 

A second reply comes from the work of Nester \emph{et al.}, according to whom different pseudotensors describe  different quasi-localizations with physical meaning tied to boundary conditions \cite{ChangNesterChen,NesterQuasi}. Thus different pseudotensors are right in different contexts.   Why should the same one be required in every context, given the close relationship between pseudotensors and boundaries?  

A third reply is that there is a best One True pseudotensor.  Perhaps it is  the Papapetrou-Belinfante pseudotensor or a higher-tech relative thereof  \cite{PetrovKatz2}. Clearly this third reply is incompatible with the second, but one can simply offer their disjunction, or even parts of each:  maybe some pseudotensors are always wrong, but others are right in one context or another.

 A fourth reply is rooted in old work of which the full import was perhaps not recognized \cite{BergmannConservation}:   nonuniqueness is not a distinct objection, but only a repeat of  coordinate dependence.  Bergmann, perhaps working with a restricted class of pseudotensors, noted that ``the totality of all conservation laws \ldots in one coordinate system is equivalent to one of them, stated in terms of all conceivable coordinate systems.'' \cite{BergmannConservation}.  The ``totality of all conservation laws''  refers to different pseudotensors. He shows  how to find the Einstein and the Landau-Lifshitz pseudotensors in his expression by choosing  $\delta x^{\sigma} = k^{\sigma}$ (where $k^{\sigma}$ is a set of  constants) or $\delta x^{\sigma} = \mathfrak{g}^{\sigma\alpha} k_{\alpha}.$  
  Especially if one has a reply to the coordinate dependence objection in terms of infinitely many energies,  reducing the nonuniqueness objection to the coordinate dependence objection  helps.


\section{Conclusion} 

Given modern progress, there seems to be no reason to regard gravitational energy realism as doomed.  
If gravitational energy is real and localized, then material + gravitational  energy-momentum is conserved: $ \partial_{\mu} (\mathfrak{T}^{\mu}_{\nu} + \mathfrak{t}^{\mu}_{\nu})=0,$  \emph{really} (not just formally), which for isolated systems can give  $E=constant.$ If space-time is  $\langle M, \eta \rangle,$ gravitational energy(s) isn't space-time energy(s) and so  do(es)n't imply substantivalism. Some classic and modern works on gravitational energy call for a reference configuration, while some work on quantum gravity argues that a (nonnegative) constant curvature background geometry must  be flat \cite{DvaliSmatrixNoDeSitter}.  Given  $\eta$, it is plausible to split $g_{\mu\nu}$ into $\gamma_{\mu\nu}$ and $\eta$. Hence there is coherence between the space-time metric split and  promising  local representations of gravitational energy. This package of views seems plausible given particle physics egalitarianism, an option traditionally rarely entertained in philosophy or conceptual discussions of  GR.

\section{Acknowledgments}

Many thanks are due to editor Antonio Vassallo and to Patrick Duerr for helpful comments that improved the paper.  All remaining errors and all opinions are my own.

%
%

\end{document}